\documentclass[twocolumn]{aastex62}

\usepackage{txfonts}
\usepackage{longtable}
\usepackage{rotating}
\usepackage{natbib}
\usepackage{graphicx}
\usepackage{graphics}
\usepackage{psfrag}
\usepackage{amssymb}
\bibpunct{(}{)}{;}{a}{}{,}

\def\Reff{$R_{\rm eff}$}
\def\Fxuv{$F_{\rm XUV}$}

\def\ergscm{$\rm erg\,cm^{-2}\,s^{-1}$}
\def\Teq{$T_{\rm eq}$}

\def\Rpl{$R_{\rm pl}$}
\def\Mpl{$M_{\rm pl}$}
\def\Re{\ensuremath{R_{\oplus}}}
\def\Me{\ensuremath{M_{\oplus}}}
\def\Mo{\ensuremath{M_{\odot}}}

\def\gs{$\rm g\,s^{-1}$}

\submitjournal{ApJL}

\shorttitle{Overcoming the energy-limited approximation for planet atmospheric escape}
\shortauthors{Kubyshkina et al.}

\begin{document}

\vspace{10 cm}
\title{Overcoming the limitations of the energy-limited approximation for planet atmospheric escape}

\correspondingauthor{Daria Kubyshkina}
\email{daria.kubyshkina@oeaw.ac.at}

\author{D. Kubyshkina}
\affiliation{Space Research Institute, Austrian Academy of Sciences, Schmiedlstrasse 6, A-8042 Graz, Austria}

\author[0000-0003-4426-9530]{L. Fossati}
\affiliation{Space Research Institute, Austrian Academy of Sciences, Schmiedlstrasse 6, A-8042 Graz, Austria}

\author{N. V. Erkaev}
\affiliation{Institute of Computational modeling of the Siberian Branch of the Russian Academy of Sciences, 660036 Krasnoyarsk, Russia}
\affiliation{Siberian Federal University, 660041 Krasnoyarsk, Russia}

\author{P. E. Cubillos}
\affiliation{Space Research Institute, Austrian Academy of Sciences, Schmiedlstrasse 6, A-8042 Graz, Austria}

\author{C. P. Johnstone}
\affiliation{Institute for Astronomy, University of Vienna, T\"urkenschanzstrasse 17, A-1180 Vienna, Austria}

\author{K. G. Kislyakova}
\affiliation{Institute for Astronomy, University of Vienna, T\"urkenschanzstrasse 17, A-1180 Vienna, Austria}
\affiliation{Space Research Institute, Austrian Academy of Sciences, Schmiedlstrasse 6, A-8042 Graz, Austria}

\author{H. Lammer}
\affiliation{Space Research Institute, Austrian Academy of Sciences, Schmiedlstrasse 6, A-8042 Graz, Austria}

\author{M. Lendl}
\affiliation{Space Research Institute, Austrian Academy of Sciences, Schmiedlstrasse 6, A-8042 Graz, Austria}

\author{P. Odert}
\affiliation{IGAM/Institute of Physics, University of Graz, Universit\"atsplatz 5, A-8010 Graz, Austria}
\affiliation{Space Research Institute, Austrian Academy of Sciences, Schmiedlstrasse 6, A-8042 Graz, Austria}








%
\begin{abstract}
Studies of planetary atmospheric composition, variability, and evolution require appropriate theoretical and numerical tools to estimate key atmospheric parameters, among which the mass-loss rate is often the most important. In evolutionary studies, it is common to use the energy-limited formula, which is attractive for its simplicity but ignores important physical effects and can be inaccurate in many cases. To overcome this problem, we consider a recently developed grid of about 7000 one-dimensional upper-atmosphere hydrodynamic models computed for a wide range of planets with hydrogen-dominated atmospheres from which we extract the mass-loss rates. The grid boundaries are [1:39]\,\Me\ in planetary mass, [1:10]\,\Re\ in planetary radius, [300:2000]\,K in equilibrium temperature, [0.4:1.3]\,\Mo\ in host star's mass, [0.002:1.3]\,au in orbital separation, and about [10$^{26}$:5$\times$10$^{30}$]\,erg\,s$^{-1}$ in stellar X-ray and extreme ultraviolet luminosity. We then derive an analytical expression for the atmospheric mass-loss rates based on a fit to the values obtained from the grid. The expression provides the mass-loss rates as a function of planetary mass, planetary radius, orbital separation, and incident stellar high-energy flux. We show that this expression is a significant improvement to the energy-limited approximation for a wide range of planets. The analytical expression presented here enables significantly more accurate planetary evolution computations without increasing computing time.
\end{abstract}
%

%
\section{Introduction} \label{sec:introduction}
The discovery of a large escaping atmosphere surrounding the hot Jupiter HD\,209458b \citep{vidal2003} stimulated a number of observational and theoretical works aimed at studying and understanding planetary atmospheric escape and its role in planetary evolution. The mass-loss rate has therefore become one of the key parameters of both observational and theoretical planetary upper-atmosphere studies.

Numerous codes have been developed over the last few years that attempt to model upper atmospheres and escape for a diversity of planets \citep[e.g.,][]{yelle2004, tian2005, murray2009, owen2012, koskinen2013a, bisikalo2013, bourrier2013, kislyakova2014, ildar2014, shematovich2014, salz2015, erkaev2016, guo2016, kubyshkina2018b, Johnstone2018}. These complex modeling tools account for a variety of physical and chemical processes describing the interaction between the planetary atmosphere and the host star's high-energy (X-ray and extreme ultraviolet (XUV)) radiation and wind. Due to their typically long computation time, these codes are more suitable to study single systems in detail, while exoplanet evolution and population models have to employ analytical approximations that are significantly faster, though less accurate.

The most widely used approximation to estimate the planetary mass-loss rate ($\dot{M}$) on the basis of the system parameters is the energy-limited equation \citep{watson1981,erkaev2007}
\begin{equation}
\dot{M}_{\rm en}=\frac{\pi \eta R_{\rm pl} R_{\rm {eff}}^2 F_{\rm XUV}}{G M_{\rm pl} K}\,, \label{eq:energyLimited}
\end{equation}
where $G$ is the universal gravitational constant, \Rpl\ is the planetary radius, \Reff\ is the effective radius at which the XUV stellar radiation is absorbed in the upper atmosphere \citep{erkaev2007,erkaev2015}, $\eta$ is the heating efficiency, $R_{\rm pl}$ is the planetary radius, $F_{\rm XUV}$ is the stellar XUV flux received by the planet, and $M_{\rm pl}$ is the planetary mass. The factor $K$ accounts for the Roche-lobe effects \citep{erkaev2007}. Equation~(\ref{eq:energyLimited}) works well for classical hot Jupiters, where the escape is hydrodynamic and driven by the stellar XUV flux \citep[e.g.,][]{lammer2003,lecavelier2007,salz2016}. However, it significantly underestimates mass loss for highly irradiated, low-density planets, where the escape is driven by a combination of the planetary intrinsic thermal energy and low gravity \citep[``boil-off''; e.g.,][]{stokl2016,lammer2016,owen2016,fossati2017}, and overestimates it for planets with hydrostatic atmospheres, where mass loss is controlled by Jeans escape \citep[e.g.,][]{salz2016,fossati2018}. In addition, Equation~(\ref{eq:energyLimited}) requires {\it a priori} knowledge of \Reff\ and $\eta$, which need complex models to be computed \citep[e.g.,][]{owen2012,shematovich2014,erkaev2015,salz2016,kubyshkina2018b}. While $\eta$ does not vary too much with system parameters \citep{salz2016} and in first approximation it is between 10\% and 20\% \citep{shematovich2014}, \Reff\ can vary significantly \citep{kubyshkina2018b}. In addition, Equation~(\ref{eq:energyLimited}) does not account for the effects of dissociation and ionization of molecular hydrogen and does not take into account the fact that in a highly supersonic atmosphere, much of the input energy ends up in the form of the kinetic energy of the gas, which is a big advantage of the hydrodynamic model.

To overcome these problems, in \citet{kubyshkina2018b} we presented a large grid of hydrodynamic upper-atmosphere models for planets less massive than 40\,\Me\ and an interpolation routine allowing the extraction of the model output parameters for planets within the grid boundaries. Here, we go a step further and present an analytical expression for the mass-loss rates as a function of system parameters developed on the basis of the grid results. By construction, this expression has the advantage over Equation~(\ref{eq:energyLimited}) of correctly accounting for \Reff\ and more adequately reproducing mass-loss rates even in cases where Equation~(\ref{eq:energyLimited}) is not applicable.

This Letter is organized as follows. Section~\ref{sec:grid} briefly presents the hydrodynamic model used to compute the grid and the grid boundaries. Section~\ref{sec:ap} gives the analytical approximation for the mass-loss rates and describes the procedure that we followed to obtain it. Section~\ref{sec:discussion} presents the discussion of our results and Section~\ref{sec:conclusions} summarizes the Letter and draws the conclusions.
\section{Hydrodynamic model and grid boundaries}\label{sec:grid}
This Letter is based on the grid of upper atmosphere models described in detail by \citet{kubyshkina2018b}. The one-dimensional hydrodynamic model employed to compute the grid describes atmospheric heating through absorption of the stellar XUV flux and accounts for hydrogen dissociation, recombination and ionization, and Ly$\alpha$- and $H_3^+$-cooling. For each model, the lower and upper boundaries are the planetary photosphere and Roche-lobe, respectively. To speed up computations, the stellar XUV spectra are reduced to two wavelengths quantifying the total extreme ultraviolet (EUV; 60\,nm) and X-ray (5\,nm) emission. All of the models have been computed assuming $\eta$\,=\,15\%.

Each model output comprises the radial profiles for the atmospheric velocity, temperature, and density of the considered species. From these, we estimated the mass-loss rate, as the outflow through the upper boundary, the effective radius \Reff, and the positions of the maximum dissociation and ionization. The grid is an ensemble of about 7000 models covering systems ranging from 1 to 39\,\Me\ in $M_{\rm pl}$, from 1 to 10\,\Re\ in $R_{\rm pl}$, from about 10$^{26}$ to 5$\times$10$^{30}$\,erg\,s$^{-1}$ in stellar XUV luminosity, from 300 to 2000\,K in planetary equilibrium temperature (\Teq), from 0.4 to 1.3\,\Mo\ in host star's mass ($M_*$), and therefore from 0.002 to 1.3\,au in orbital separation, which is not an independent parameter of the model, but derived from $M_*$ and \Teq \citep[see][for more details]{kubyshkina2018b}. To avoid computing probably unphysical planets, we limited the computations to planets with an average density larger than 0.03\,$g\,cm^{-3}$, a Roche radius larger than 1.5\,\Rpl, and a restricted Jeans escape parameter $\Lambda$ smaller than 80 \citep{fossati2017}, where
\begin{equation}\label{eq:lambda}
\Lambda = \frac{G M_{\rm pl} m_{\rm H}}{k_{\rm b} T_{\rm eq} R_{\rm pl}}\,,
\end{equation}
is the value of the Jeans escape parameter \citep{jeans1925,chamber1963,opik1963} calculated at the observed planetary radius and mass for the planet's \Teq\ and considering atomic hydrogen, where $m_{\rm H}$ is the mass of the hydrogen atom and $k_{\rm b}$ is the Boltzmann constant.
\section{Analytical formulation of the mass-loss rates}\label{sec:ap}
We took each point available in the grid to obtain an analytical approximation for the mass-loss rates as a function of $\Lambda$, $R_{\rm pl}$, orbital separation ($d_0$), and $F_{\rm XUV}$. The equilibrium temperature is not taken into account as an input parameter, because the stellar radius varies weakly over the main-sequence lifetime of a star, thus \Teq\ depends almost exclusively on the orbital separation \citep[see][for more details]{kubyshkina2018b}. Furthermore, we do not account for the stellar mass as an input parameter, because for the vast majority of the cases considered here its effect on the results is significantly smaller than the difference between the approximated and modeled mass-loss rates.

We start by considering that the mass-loss rates as a function of $\Lambda$ can be written as \citep{kubyshkina2018b}
\begin{equation}
\label{eqn:ap0}
\ln(\dot{M}) = C + K\,\ln(\Lambda)\,,
\end{equation}
where $C$ and $K$ are coefficients that depend on the system parameters. The term $C$ defines the maximum level of hydrodynamic escape at $\Lambda$\,=\,1. Equation~(\ref{eqn:ap0}) can be rewritten as
\begin{equation}
\dot{M} = \tilde{C}\,\,\Lambda^K\,,
\end{equation}
where $\tilde{C}$\,=\,$e^C$. Assuming that \Fxuv, $d_0$, and \Rpl\ are independent parameters of the function $\tilde{C}$ one can write
\begin{equation}
\tilde{C} = f(F_{\rm XUV})\,\,g(d_0)\,\,h(R_{\rm pl})\,,
\end{equation}
where $f$, $g$, and $h$ are functions of one variable. It therefore follows that
\begin{equation}
\label{eqn:ap1}
C = \ln(f(F_{\rm XUV})) + \ln(g(d_0)) + \ln(h(R_{\rm pl}))\,.
\end{equation}
From the distribution of the mass-loss rates in the grid as a function of input parameters, we concluded that $f$, $g$, and $h$ can be best approximated by power laws of the form $\beta_i\,x^{\alpha_i}$, where $x$ is one of \Fxuv, ${d_0}/{\rm au}$, or ${R_{\rm pl}}/{R_{\oplus}}$. Finally, Equation~(\ref{eqn:ap1}) takes the form
\begin{equation}
\label{eq:Clong} C = \beta + \alpha_1\,\ln(\frac{F_{\rm XUV}}{erg\,s^{-1}cm^{-2}}) + \alpha_2\,\ln(\frac{d_0}{\rm AU}) + \alpha_3\,\ln(\frac{R_{\rm pl}}{R_{\oplus}})\,,
\end{equation}
where $\beta = \sum ln(\beta_i)$.

The term $K$ in Equation~(\ref{eqn:ap0}) describes how fast the mass-loss rates decrease with increasing $\Lambda$. From the results of the grid, we noticed that $K$ depends on the orbital separation (i.e., $K = K(d_0)$), which we express as
\begin{equation}
\label{eq:Klong} K = \zeta + \theta\,\,\ln(\frac{d_0}{\rm AU})\,,
\end{equation}
where $\theta$ and $\zeta$ also depend on the system parameters. For the other input parameters $\theta$ is below $10^{-4}$, which is why we ignore how $K$ depends on them.

By combining Equations~(\ref{eqn:ap0}), (\ref{eq:Clong}), and (\ref{eq:Klong}) one arrives at the final form for the analytical expression of the mass-loss rates in \gs\ as a function of the input parameters, which we call ``hydro-based approximation,'' as
\begin{equation}
\label{eqn:ap2} \dot{M}_{\rm HBA} = e^{\beta}\,\,(F_{\rm XUV})^{\alpha_1}\,\,\left(\frac{d_0}{\rm AU}\right)^{\alpha_2}\,\,\left(\frac{R_{\rm pl}}{R_{\oplus}}\right)^{\alpha_3}\,\,\Lambda^{K}\,.
\end{equation}
where $\beta$, $\alpha_1$, $\alpha_2$, $\alpha_3$, $\zeta$, and $\theta$ are the coefficients listed in Table~\ref{tab:coefficients} that are different depending on whether a planet has a $\Lambda$ value greater or smaller than $e^{\Sigma}$, where
\begin{equation}
\label{eq:sigma} \Sigma = \frac{15.611 - 0.578\,\ln(F_{\rm XUV}) + 1.537\,\ln(\frac{d_0}{\rm AU}) + 1.018\,\ln(\frac{R_{\rm pl}}{R_{\oplus}})}{5.564 + 0.894\,\ln(\frac{d_0}{\rm AU})}\,.
\end{equation}

\begin{figure}
\includegraphics[width=\hsize]{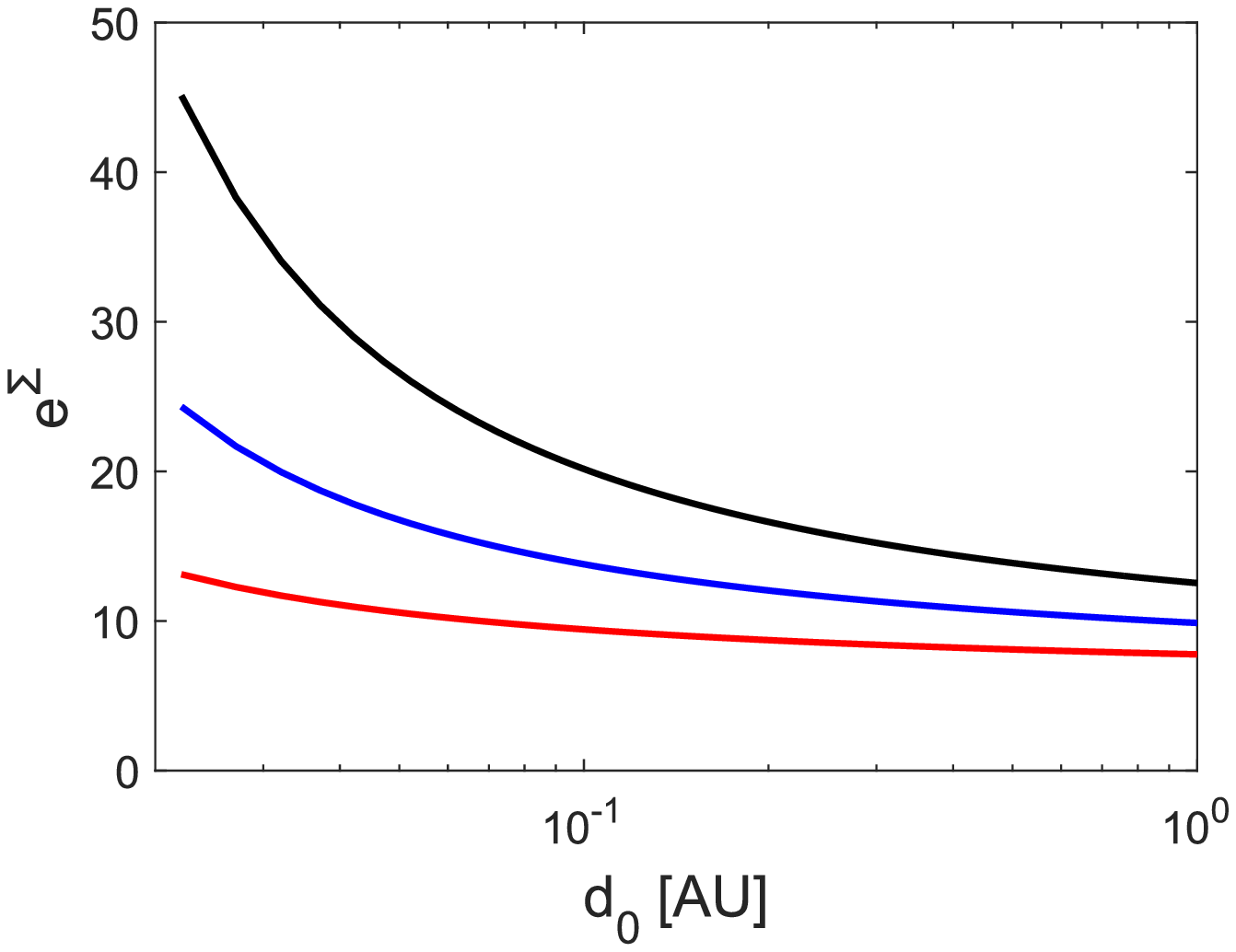}\\
\includegraphics[width=\hsize]{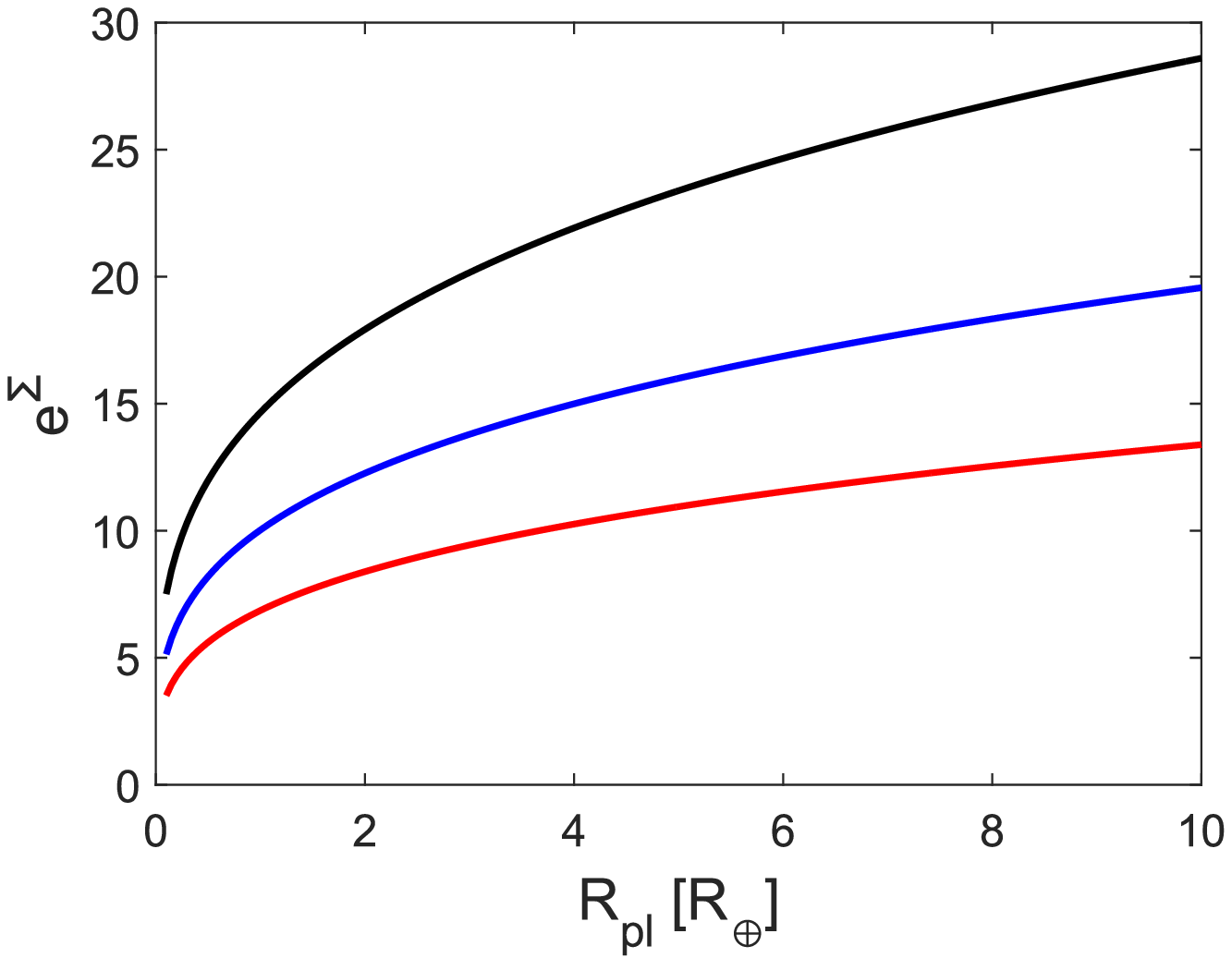}
\caption{Value of $e^{\Sigma}$ as a function of planetary radius at a fixed orbital separation of 0.1\,au (top panel) and as a function of orbital separation at a fixed planetary radius of 3\,\Re (bottom panel) for \Fxuv\ values of 100 (black), 1000 (blue), and 10,000\,\ergscm (red).} \label{fig:applim}
\end{figure}

The different behavior of the approximation for the different values of $\Lambda$ is connected to the fact that at small $\Lambda$ values (i.e., below $e^{\Sigma}$) the main driver of atmospheric escape is a combination of the planetary intrinsic thermal energy and low gravity. This leads to a significant change in the dependence of the mass-loss rates on the system parameters. Figure~\ref{fig:applim} shows the shape of the boundary defined by $e^{\Sigma}$ as a function of planetary radius and orbital separation for three \Fxuv\ values.
\begin{table}
\caption{Parameters of the hydro-based approximation present in Equations~(\ref{eq:Klong}) and (\ref{eqn:ap2}) obtained using iterative least squares estimation.}
\label{tab:coefficients}
\begin{tabular}{c|c|c|c|c|c|c}
\hline
\hline
 & $\beta$ & $\alpha_1$ & $\alpha_2$ & $\alpha_3$ & $\zeta$ & $\theta$ \\
\hline
$\Lambda\,<\,e^{\Sigma}$ & 32.0199 & 0.4222 & -1.7489 & 3.7679 & -6.8618 & 0.0095 \\
$\Lambda\,\geq\,e^{\Sigma}$ & 16.4084 & 1.0000 & -3.2861 & 2.7500 & -1.2978 & 0.8846 \\
\hline
\end{tabular}
\end{table}

\begin{figure*}
\includegraphics[width=0.5\hsize]{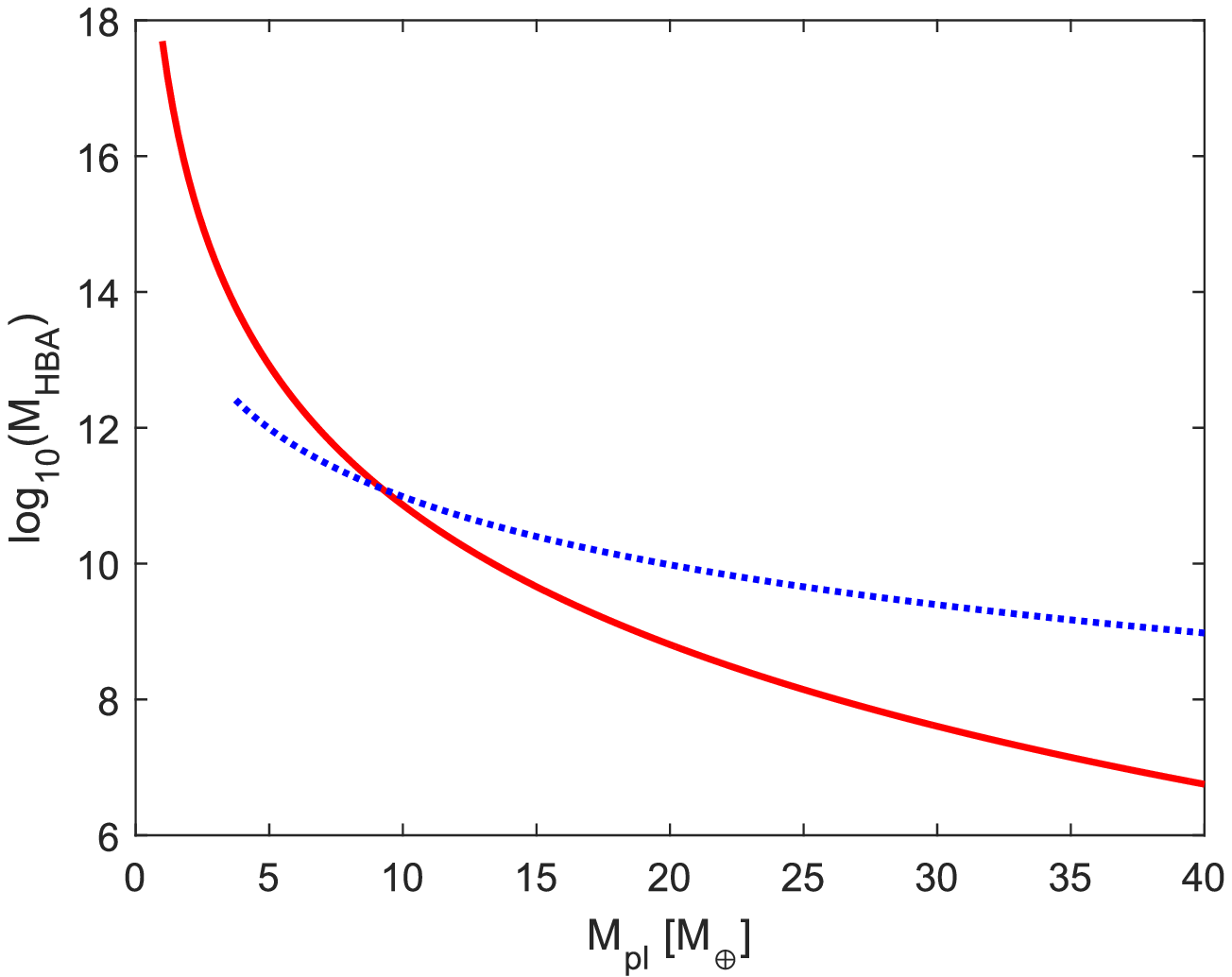}
\includegraphics[width=0.5\hsize]{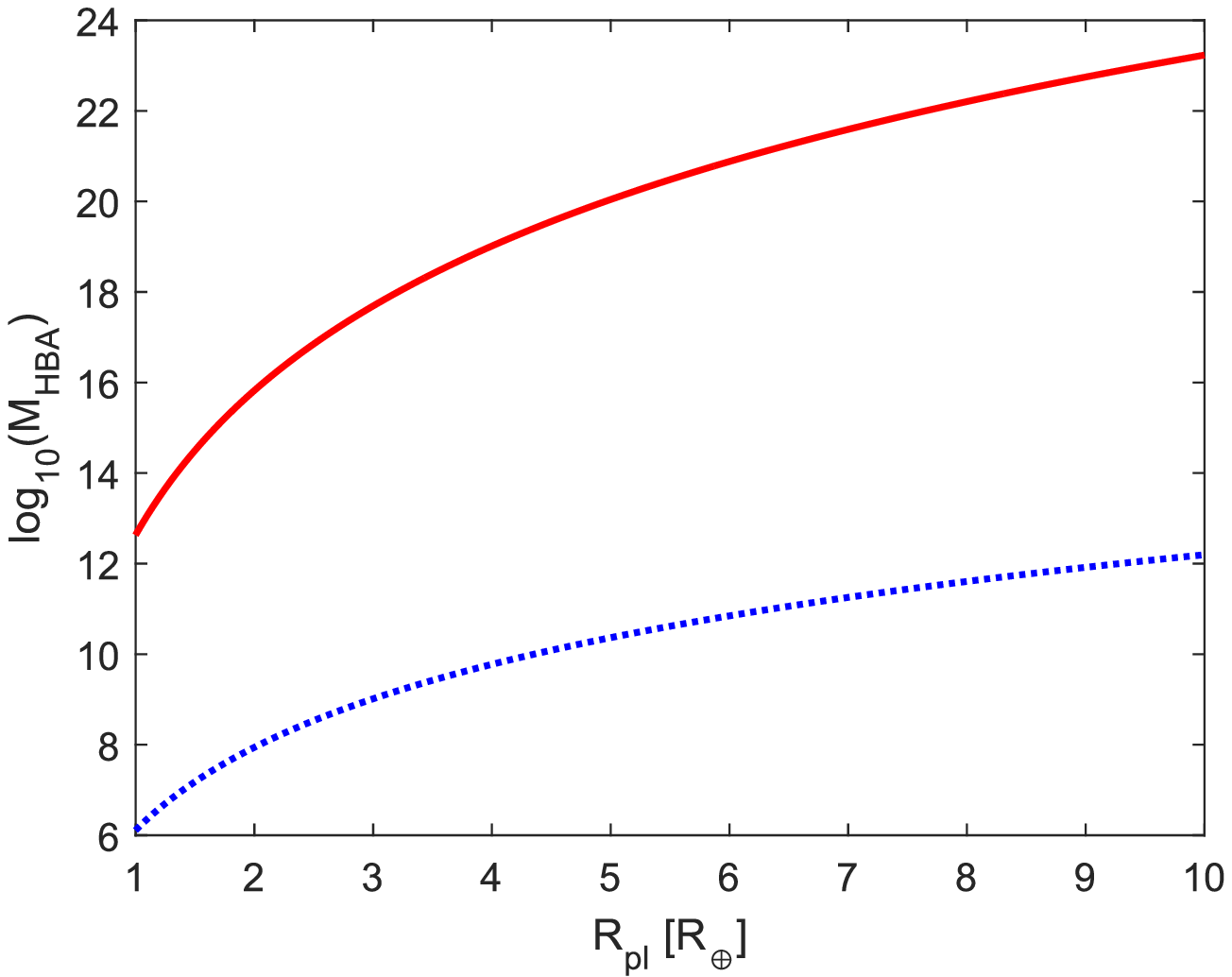} \includegraphics[width=0.5\hsize]{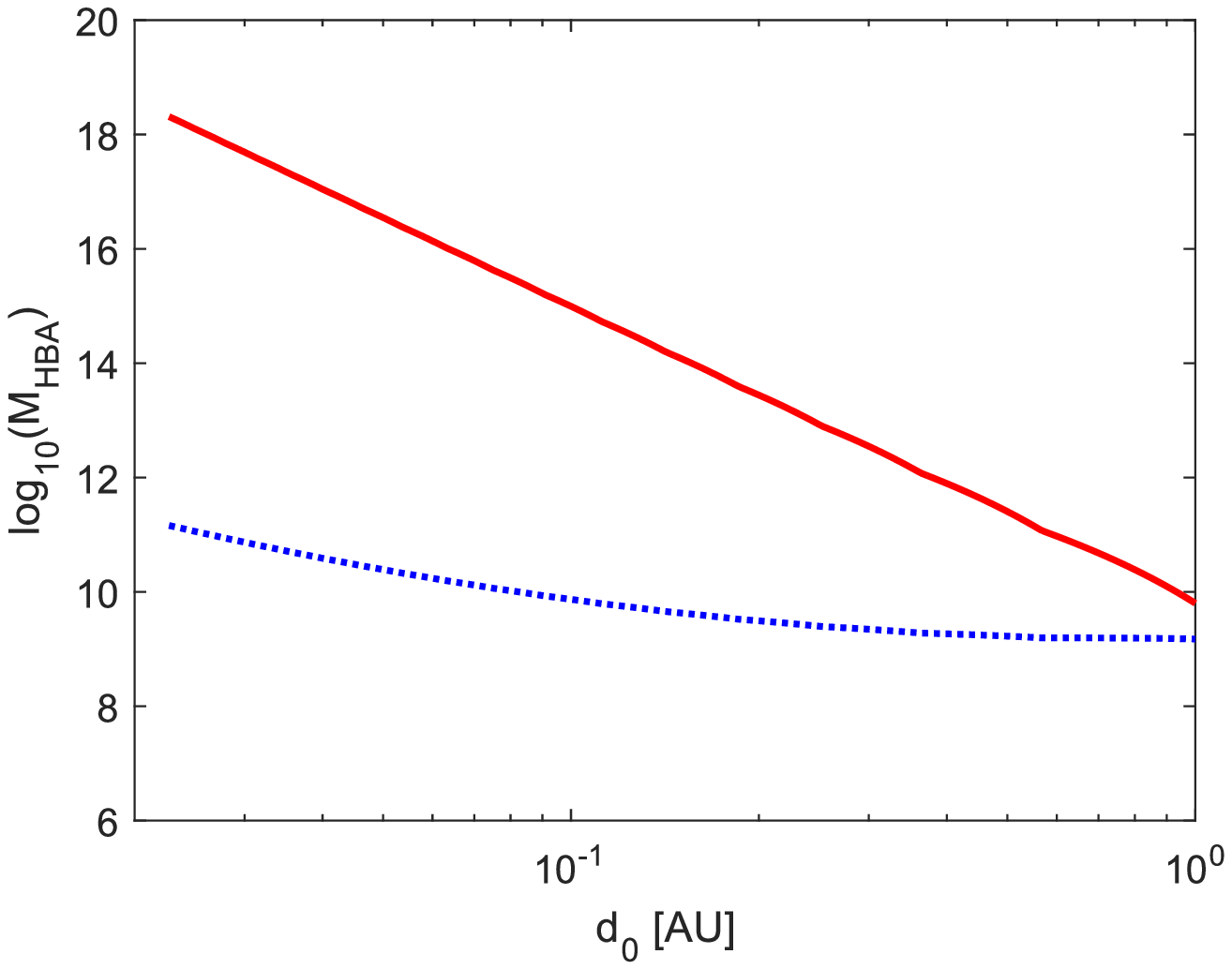} \includegraphics[width=0.5\hsize]{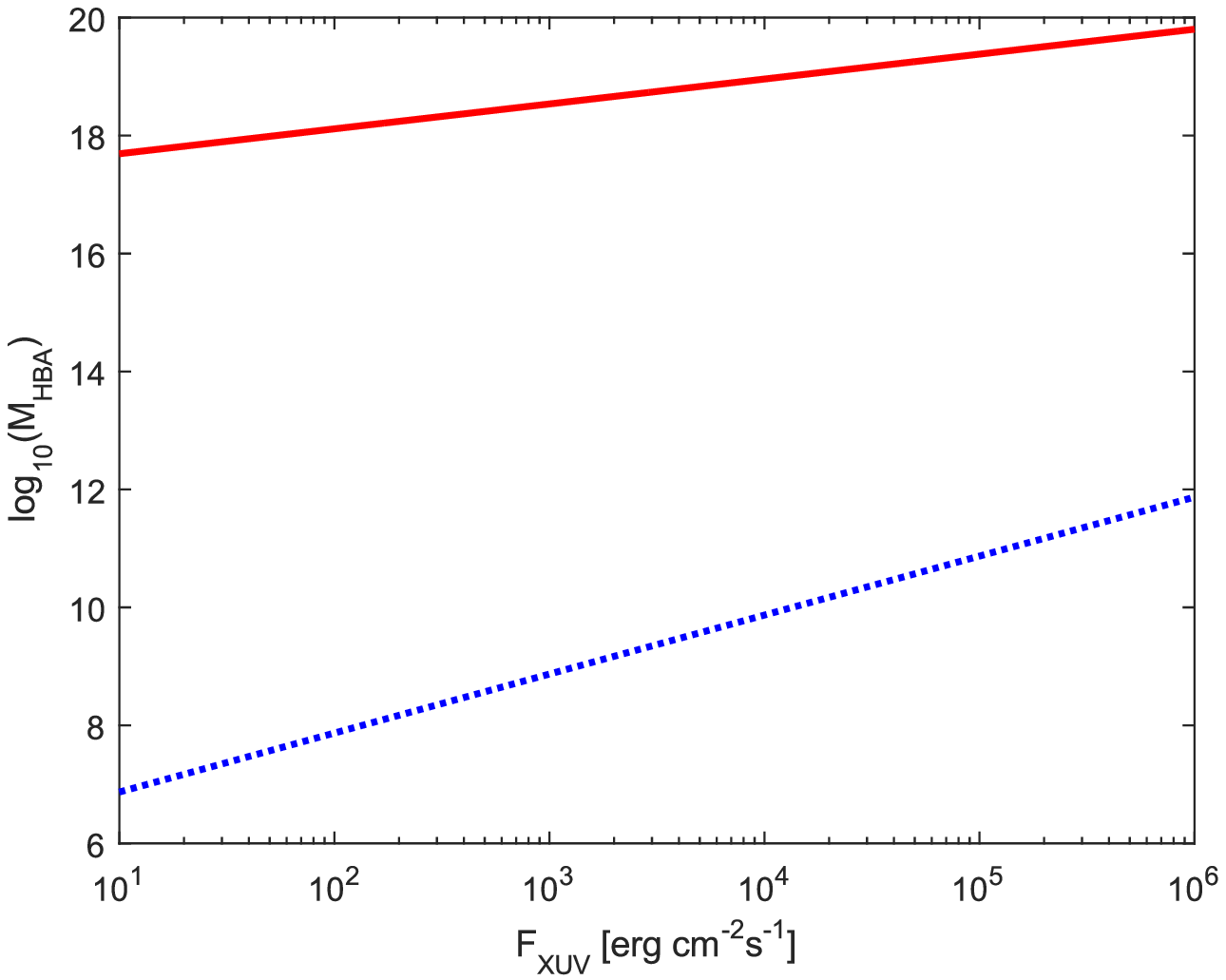}
\caption{Behavior of the hydro-based approximation as a function of \Mpl, \Rpl, $d_0$, and \Fxuv\ for two planets, one on each side of the boundary defined by $e^{\Sigma}$. The red line is for the planet with $\Lambda$ smaller than $e^{\Sigma}$ (\Mpl\,=\,1\,\Me, \Rpl\,=\,3.0\,\Re, $d_0$\,=\,0.03\,au, and \Fxuv\,=\,10\,\ergscm), while the blue dashed line is for the planet with higher $\Lambda$ (\Mpl\,=\,39\,\Me, \Rpl\,=\,3.0\,\Re, $d_0$\,=\,0.1\,au, and \Fxuv\,=\,10,000\,\ergscm). Both planets orbit a Sun-like star (i.e., $M_*$\,=\,1\,\Mo). In the top-left panel, the blue line does not go all the way down to 1\,\Me\ because below 5\,\Me\ the planet crosses the $e^{\Sigma}$ boundary.} 
\label{fig:app_len}
\end{figure*}

Figure~\ref{fig:app_len} illustrates the behaviour of the hydro-based approximation as a function of the input parameters considering two hypothetical planets lying on different sides of the $e^{\Sigma}$ boundary. The planet with $\Lambda$ smaller than $e^{\Sigma}$ has a mass of 1\,\Me, a radius of 3.0\,\Re, orbits the 1\,\Mo\ host star at a distance of 0.03\,au, and is irradiated by an XUV flux of 10\,\ergscm. The planet with larger $\Lambda$ has a mass of 39\,\Me, a radius of 3.0\,\Re, orbits the 1\,\Mo\ host star at a distance of 0.1\,au, and is subject to an XUV flux irradiation of 10,000\,\ergscm.

The top-left panel of Figure~\ref{fig:app_len} is particularly telling. At small planetary mass, the mass-loss rates are mostly driven by the combination of high \Teq\ and low gravity. This implies that the mass-loss rates are higher for the planet with $\Lambda$ smaller than $e^{\Sigma}$ because its closer distance to the star implies a higher equilibrium temperature. As \Mpl\ increases, the mass-loss rates of both planets decrease, but at different rates such that at about 8\,\Me\ the planet with $\Lambda$ larger than $e^{\Sigma}$ has the larger mass-loss rate of the two. This is because with increasing mass, the mass-loss rates become progressively more controlled by the stellar XUV flux, which is higher for the planet with $\Lambda$ larger than $e^{\Sigma}$.

We remark that the hydro-based approximation presented in this Letter is applicable only to planets with hydrogen-dominated atmospheres. Depending on their composition, atmospheres dominated by gases other than hydrogen can react in a variety of different ways to similar levels of XUV irradiation \citep[see, e.g.][]{Johnstone2018}.
\section{Discussion}\label{sec:discussion}
\begin{figure*}[ht!]
\includegraphics[width=0.5\hsize]{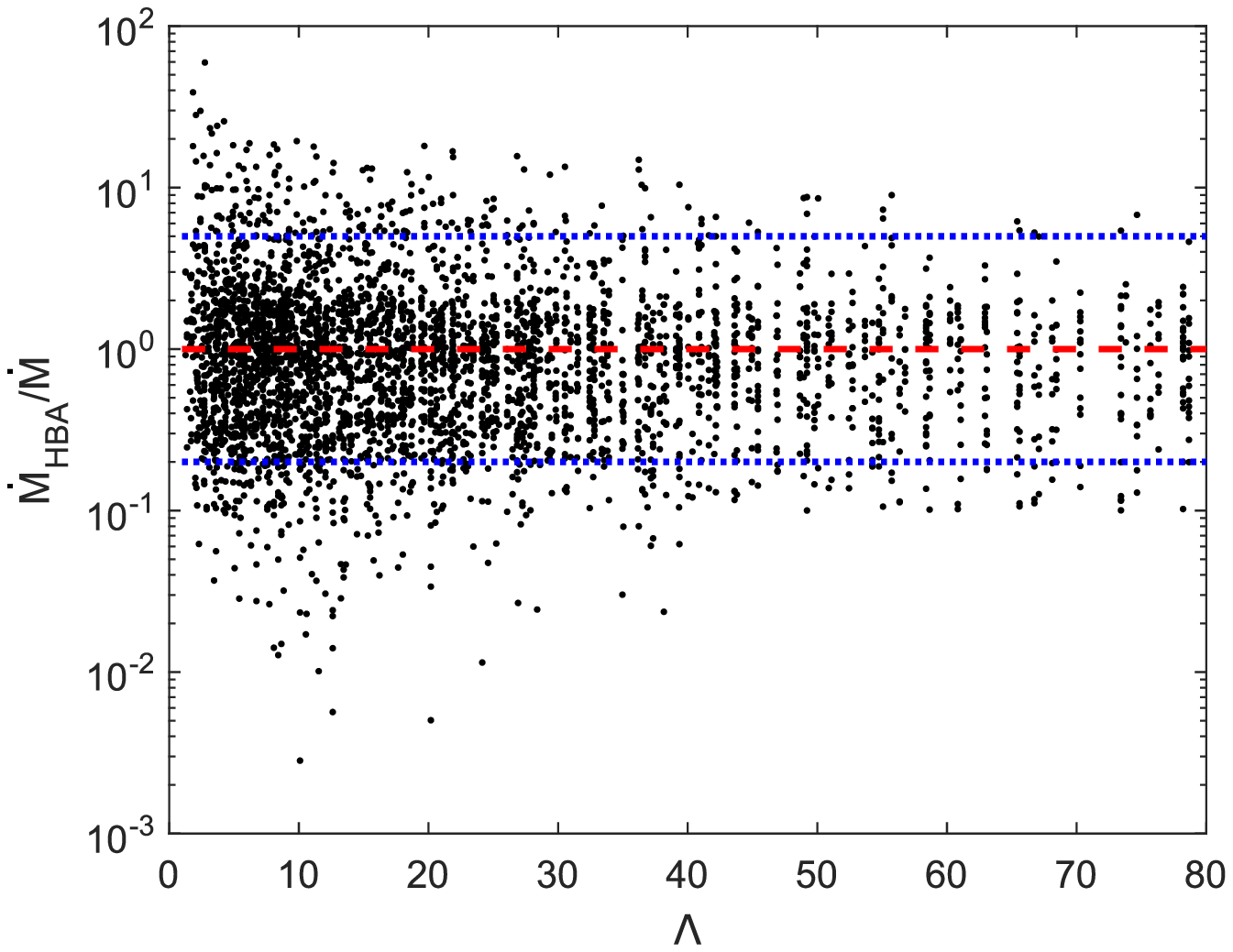}
\includegraphics[width=0.5\hsize]{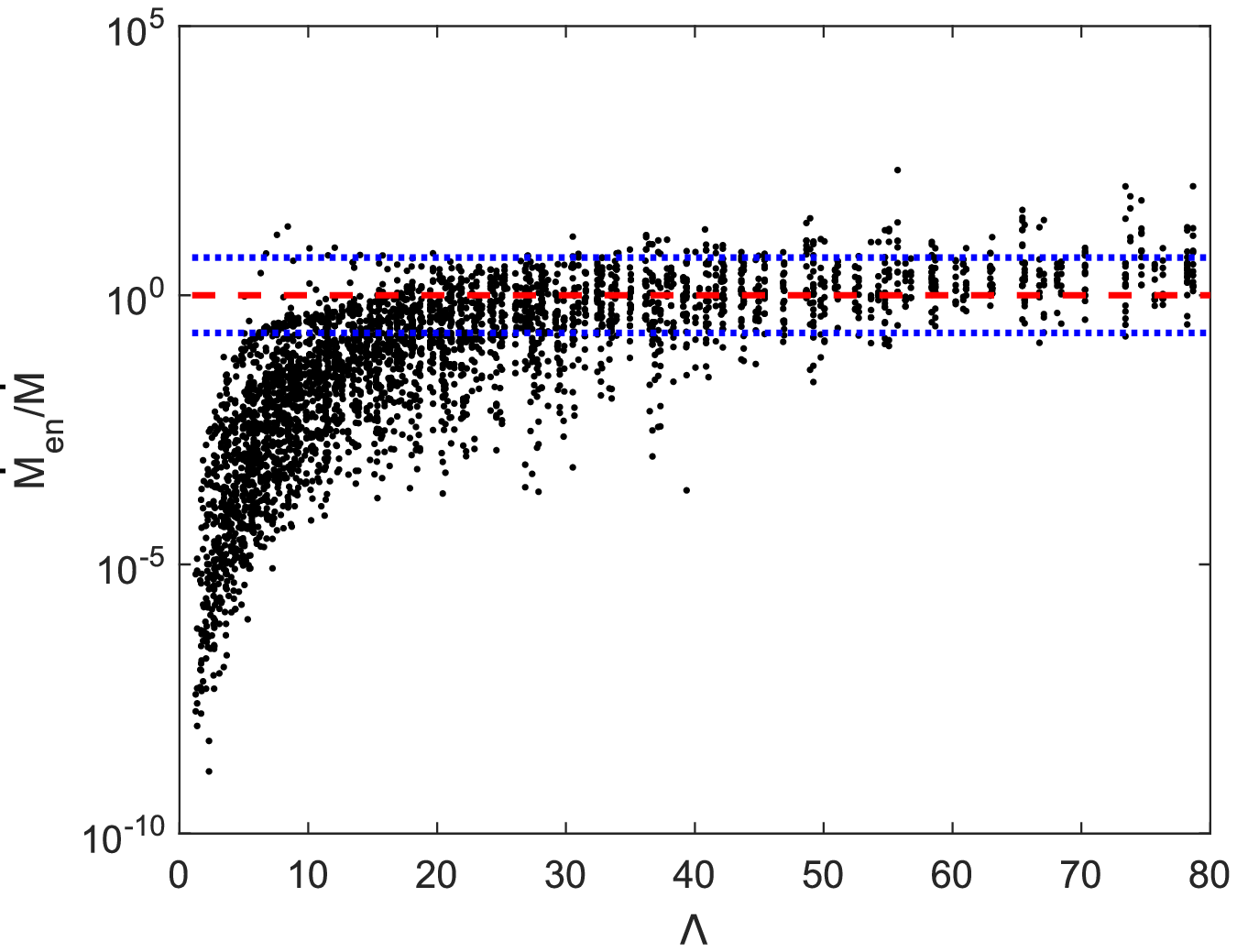}
\caption{Left panel: ratio between the mass-loss rates obtained from the hydro-based approximation ($\dot{M}_{\rm HBA}$) and from the hydrodynamic grid as a function of $\Lambda$. Right panel: the same as the left panel, but for the mass-loss rates derived from the energy-limited formula considering the \Reff\ values derived from the grid. In both panels, the red line is at one, while the blue lines are at values of 5 and 0.2. Note the large difference in the scale of the y-axis between the two plots.} 
\label{fig:ap_error}
\end{figure*}

The left panel of Figure~\ref{fig:ap_error} shows the ratio between the mass-loss rates obtained from the hydro-based approximation and from the hydrodynamic grid as a function of $\Lambda$. The vast majority of the data points cluster around unity and, in 85\% of the cases (97\% for planets with $\Lambda$ larger than 30), the hydro-based approximation deviates less than a factor of five from what is given by the grid. We also find that the accuracy of the hydro-based approximation increases with \Fxuv, and for \Fxuv\ grater than 10,000\,\ergscm\ in 90\% of the cases the deviation remains within a factor of two. The largest deviations, of the order of 10$^2$, are found for planets with the smallest $\Lambda$ and $d_0$ values. For these systems the dependence on the input parameters deviates from the considered power laws. This is probably due to the fact that for these planets the position of the Roche-lobe, thus the dependence on the stellar mass, begins to play a role, which is not considered in the hydro-based approximation.

The right panel of Figure~\ref{fig:ap_error} shows the deviation of the mass-loss rates derived from the energy-limited formula (Equation\,\ref{eq:energyLimited}) with those obtained from the grid of hydrodynamic models as a function of $\Lambda$. For the energy-limited mass-loss rates we considered a heating efficiency of 15\% (as for the computation of the grid), Roche-lobe effects, and the \Reff\ value given by the grid. At small $\Lambda$, Equation~(\ref{eq:energyLimited}) underestimates the mass-loss rates by several orders of magnitude, up to 10$^8$. This is because the energy-limited approximation does not account for escape driven by a combination of the intrinsic planetary thermal energy and low gravity. In addition, mostly for planets with intermediate $\Lambda$ values (10--30) the use of the \Reff\ value taken from the grid alleviates the discrepancy from the hydrodynamic mass-loss rates shown in the bottom panel of Fig.~\ref{fig:ap_error}. In fact, by computing the energy-limited mass-loss rates employing \Reff\ values equal to the planetary radii, the deviation would increase by up to a factor of 10--20. The energy-limited formula leads also to a slight systematic overestimation of the mass-loss rates up to a factor of 50 for planets with large $\Lambda$ values.

Table~\ref{tab:comparison} compares the mass-loss rates obtained from the hydro-based approximation, from direct computations of the hydrodynamic code described in Section~\ref{sec:grid}, from Equation~(\ref{eq:energyLimited}), and from the literature. The mass-loss rates derived from the energy-limited approximation given in Table~\ref{tab:comparison} were computed considering the \Reff\ values derived from the output of the hydrodynamic code as well as \Reff\ equal to the planetary radius (in parenthesis). This last case corresponds to the most common one when resourcing to the energy-limited approximation, namely that of no availability or possibility to use a hydrodynamic code. This comparison further shows that the hydro-based approximation is a significant improvement in comparison to the energy-limited formula (e.g., Kepler-11\,b, 55\,Cnc\,e), particularly when $\dot M_{\rm en}$ assumes $R_{\rm eff} = R_{\rm pl}$.

Table~\ref{tab:comparison} considers also the hot Jupiters HD\,209458\,b and HD\,189733\,b, which lie outside of our grid boundaries by size and therefore, in principle, also outside of the regime of validity of the hydro-based approximation. Despite this, the results shown in Table~\ref{tab:comparison} suggest that the hydro-based approximation performs well also for close-in planets outside of the upper mass-radius boundaries of the grid. This is not surprising given that for these close-in, massive planets Equation~(\ref{eq:energyLimited}) is a good approximation for the mass-loss rates and that the hydro-based approximation for planets with large $\Lambda$ values appears to work as good as the energy-limited formula, if not better. Despite this, we suggest not relying on the hydro-based approximation outside of the boundaries given by the grid on which it is based.
\begin{table*}
\caption{Comparison between the mass-loss rates obtained from the hydro-based approximation ($\dot{M}_{\rm HBA}$), from direct modeling ($\dot{M}$), from the energy-limited formula ($\dot{M}_{\rm en}$), and from the literature. The latter is further split into published measurements ($\dot{M}_{\rm publ,mea}$) and model estimates ($\dot{M}_{\rm publ,mod}$). The last two columns list also the source of the published mass-loss rates. The energy-limited mass-loss rates are listed considering the \Reff\ value computed from the hydrodynamic code and, in parenthesis, for an \Reff\ value equal to the planetary radius.} 
\label{tab:comparison}
\begin{tabular}{l|c|c|c|c|c|c|c|c|l|l}
\hline
\hline
ID & $\Lambda$ & $R_{\rm pl}$ & $d_0$ & \Fxuv & $M_*$ & $\dot{M}_{\rm HBA}$ & $\dot{M}$ & $\dot{M}_{\rm en}$ & $\dot{M}_{\rm publ,mea}$ & $\dot{M}_{\rm publ,mod}$ \\
   &        & (\Re)  & (au)  & (\ergscm)     & (\Mo) & (\gs)     & (\gs)              & (\gs)   & (\gs)               \\
\hline
HD\,209458\,b &   90 & 15.45 & 0.047   & 1086  & 1.148 &  9.6$\times10^{9}$ & 1.2$\times10^{10}$ &  8.0$\times 10^{9}$   &      2$\times 10^{10}$ (a)      & 3.3$\times 10^{10}$ (b) \\
            &      &       &         &       &       &                    &                    & (5.5$\times 10^{9}$)  & $0.6-10\times 10^{10}$ (c)      & $1.9\times 10^{10}$ (d) \\
\hline
GJ\,436\,b    &   58 &  4.25 & 0.02887 & 1760  & 0.452 &  2.3$\times10^{9}$ & 3.95$\times10^{9}$ &  2.9$\times10^{9}$    & 1$\times 10^8-1\times 10^9$ (e) & 1$\times 10^{10}$ (f) \\
            &      &       &         &       &       &                    &                    & (1.9$\times 10^{9}$)  & 2.2$\times 10^{10}$ (g)         & 4.5$\times 10^9$ (d)  \\
\hline
Kepler-11\,b  &  18  &  1.97 & 0.091   & 278   & 0.95  &  1.9$\times10^{9}$ &  1.2$\times10^{9}$ &  7.5$\times 10^{8}$   &                                 & $1.15-2\times 10^8$ (h)\\
            &      &       &         &       &       &                    &                    & (1.5$\times 10^{8}$)  &                                 & $1\times 10^9$ (f) \\
\hline
HD\,189733\,b &  179 & 12.74 & 0.03    & 24778 & 0.8   &  4.5$\times10^{9}$ &  4.9$\times10^{9}$ &  $4.8\times 10^{10}$  & $0.04-10\times 10^{10}$ (c)     & $5-9\times 10^{11}$ (f) \\
            &      &       &         &       &       &                    &                    & (4.3$\times 10^{10}$) &                                 & $4.1\times 10^9$ (d)\\
\hline
GJ\,3470\,b   &   37 &  4.18 & 0.03557 & 1868  & 0.539 & 1.6$\times10^{10}$ & 1.3$\times10^{10}$ &  $7.0\times 10^{9}$   &                                 & $4.6\times 10^{10}$ (d) \\
            &      &       &         &       &       &                    &                    & (3.0$\times 10^{9}$)  &                                 & \\
\hline
HD\,149026\,b &   61 &  8.04 & 0.04288 & 6886  & 1.3   & 4.5$\times10^{10}$ & 3.4$\times10^{10}$ &  $1.5\times 10^{10}$  &                                 & $2.7\times 10^{10}$ (d) \\
            &      &       &         &       &       &                    &                    & (9.7$\times 10^{9}$)  &                                 & \\
\hline
HAT-P-11\,b   & 48.5 &  4.72 & 0.053   & 3236  & 0.81  & 1.3$\times10^{10}$ & 1.1$\times10^{10}$ &  $6.8\times 10^{9}$   &                                 & $1.9\times 10^{10}$ (d) \\
            &      &       &         &       &       &                    &                    & (4.0$\times 10^{9}$)  &                                 & \\
\hline
55\,Cnc\,e    &   16 &  1.99 & 0.01544 & 570   & 0.905 & 4.9$\times10^{10}$ & 4.2$\times10^{10}$ &  $9.0\times 10^{8}$   & $3.0\times 10^8$ (g)            & $1.4\times 10^{10}$ (d) \\
            &      &       &         &       &       &                    &                    & (1.7$\times 10^{8}$)  &                                 & \\
\hline
HD\,97658\,b  &   34 &  2.24 & 0.08    & 955   & 0.85  &  1.8$\times10^{9}$ &  1.7$\times10^{9}$ &  $1.0\times 10^{9}$   &                                 & $3.0\times 10^{9}$ (d) \\
            &      &       &         &       &       &                    &                    & (4.4$\times 10^{8}$)  &                                 & \\
\hline
\end{tabular}
\tablecomments{ a -- \citet{ehrenreich2008}; b --
\citet{murray2009}; c -- \citet{bourrier2013}; d --
\citet{salz2016}; e -- \citet{ehrenreich2015}; f --
\citet{guo2016}; g -- \citet{bourrier2016}; h --
\citet{lammer2013}.}
\end{table*}
%
\section{Conclusions}\label{sec:conclusions}
We present here an analytical approximation, called hydro-based approximation, for the mass-loss rates of planets between 1 and 40\,\Me\ hosting a hydrogen-dominated atmosphere. The aim is to overcome the limits of the widely used energy-limited approximation. The hydro-based approximation is based on a grid of almost 7000 one-dimensional hydrodynamic upper-atmosphere models covering systems ranging from 1 to 39\,\Me\ in $M_{\rm pl}$, 1 to 10\,\Re\ in $R_{\rm pl}$, 300 to 2000\,K in planetary equilibrium temperature, 0.4--1.3\,\Mo\ in host star's mass, 0.002 to 1.3\,au in orbital separation, and about 10$^{26}$ to 5$\times$10$^{30}$\,erg\,s$^{-1}$ in stellar XUV luminosity. These boundaries describe also the range of validity of the hydro-based approximation presented here.

By construction, the hydro-based approximation is a much better representation of the mass-loss rates derived from the hydrodynamic code compared to the energy-limited approximation. In particular, for most planets with small $\Lambda$ values (i.e., $\lesssim$20), the hydro-based approximation outperforms the energy-limited approximation by several orders of magnitude. This is the regime in which atmospheric escape is the strongest and is driven by a combination of the high intrinsic planetary thermal energy and low gravity.

The hydro-based approximation has a further important advantage over the energy-limited approximation: it does not require {\it a priori} knowledge of \Reff, the exact value of which can be obtained only from hydrodynamical computations; this would, in turn, make the energy-limited approximation redundant. These arguments clearly demonstrate the significant improvement of the hydro-based approximation over what is given by the energy-limited formula.

Being fully analytical, the hydro-based approximation can be employed in planetary evolution computations without significantly increasing the computing time. Furthermore, the hydro-based approximation produces significantly more adequate results than the energy-limited approximation exactly for those planets for which atmospheric escape is most significant.

We are continuing to enlarge the grid of hydrodynamic models on which the hydro-based approximation is based. This will enable us in the near future to extend the approximation also to other kinds of planets (with higher masses and densities), thus making the hydro-based approximation a key resource for the computation of planetary population and evolution models. Such advances are necessary to identify the role of atmospheric escape in shaping the observed exoplanet population, particularly at a time in which space missions such as {\it TESS}, {\it ARIEL} and {\it PLATO} will detect and measure the basic parameters of thousands of nearby planetary systems.

%
\acknowledgments
We acknowledge the FFG project P853993, the FWF/NFN projects S11607-N16, S11604-N16, and the FWF projects P27256-N27 and P30949-N36. N.V.E. acknowledges support by RFBR grant No. 18-05-00195-a and 16-52-14006 ANF\_a. We thank the anonymous referee for useful comments.

%

\vspace{5mm}

\end{document}